\documentclass[preprint,aps,showpacs]{revtex4}
\usepackage{epsfig}
\usepackage{graphics}
\usepackage{graphicx}

\newcommand{\beq}{\begin{equation}}               
\newcommand{\eeq}{\end{equation}}                 
\newcommand{\bqry}{\begin{eqnarray}}              
\newcommand{\eqry}{\end{eqnarray}}                
\newcommand{\bqryn}{\begin{eqnarray*}}            
\newcommand{\eqryn}{\end{eqnarray*}}              

\begin{document}

\title{\bf Pion LINAC as an Energy-Tagged $\nu$ Source}
\author{T.\ Goldman}\email{ tgoldman@lanl.gov}
\affiliation{Theoretical Division, MS-B283, 
Los Alamos National Laboratory, Los Alamos, NM 87545}
\author{Richard R.\ Silbar}\email
{ silbar@lanl.gov} 
\affiliation{Theoretical Division, MS-B283, 
Los Alamos National Laboratory, Los Alamos, NM 87545}

\begin{flushright}
Revised Version\\
January 14, 2009\\
{LA-UR-08-04369}\\
{arXiv:0807.0461}\\
\end{flushright}

\begin{abstract}

The energy spectrum and flux of neutrinos from a 
linear pion accelerator are calculated analytically 
under the assumption of a uniform accelerating gradient.
The energy of a neutrino from this source reacting in a detector 
can be determined from timing and event position information.

\end{abstract} 


\maketitle

\section{Introduction}

There has been considerable analysis \cite{FNAL} of the value
of a high energy muon storage ring for the purpose of producing
well-defined, high energy beams of neutrinos. This approach takes
advantage of the relatively long muon lifetime to allow for collection
at modest energies and acceleration to very high energies, where the
decay neutrinos form an intense, highly collimated beam. The proposals
for such systems are quite expensive since the (generally hadronically)
produced muons must be captured, cooled and accelerated, all on a
microsecond time scale.

Some time ago, Los Alamos proposed  \cite{PILAC} a pion post-accelerator
(PILAC) to be appended to the then extant meson factory, LAMPF. The 
proposal took advantage of the high pion flux produced at modest energies, 
to collimate and inject into a high gradient linac. Phase rotation was
introduced to reduce the energy spread of the beam considerably.  The
main practical difficulty with this relatively modest cost proposal was
the short pion lifetime -- 26 nanoseconds. This required a very high
accelerating gradient in order to achieve a useable flux at the exit of
the acceleration section.

The decay of a positively-charged pion is almost exclusively into anti-muons and 
muon-neutrinos (or muons and muon-antineutrinos for negatively-charged pions) 
with a purity at the part in ten thousand level.
Thus a PILAC also provides an intense, collimated beam of neutrinos with
an energy spectrum ranging from negligible to initially half the pion input 
energy and finally to half the output energy value. We show here that the 
neutrino spectrum, $d N_\nu/d E_\nu$, can adjusted to be almost flat in energy, 
if sufficiently high accelerating gradients for the pions can be achieved. 
The necessary values are below those already achieved \cite{LEP} using
superconducting RF cavities.\cite{ftn1} 

However, there is another advantage to such a neutrino source that does not 
dependent upon whether or not high gradients can be achieved. Currently, it 
is very difficult to determine the energy of the neutrino involved in a particular 
scattering event in the detector.  Usually, a complicated series of Monte Carlo 
simulations of the detector and its response defines a statistical determination 
only.  In the scheme we are discussing, however, it is possible to obtain the 
incident neutrino energy {\em directly, event by event}.

We analyze here the theoretically simple case of a uniform accelerating 
gradient analytically.  A more realistic case with drift sections is left for the 
accelerator physicists who would actually design and build such a system. 

\section{Calculations}

Although we ultimately want to obtain the neutrino energy 
spectrum as a function of distance from the injection point 
of the accelerator, to follow the decay rate, it turns out 
to be more convenient to first find the pion energy and 
decay rate as functions of time.  Throughout, we will use 
units where $\hbar = c = 1$. 

\subsection{Time Dependence}

We begin with pions injected at energy $E_{I}$ and $\gamma_{I} 
\equiv E_{I}/m_\pi$ where $m_\pi$ is the pion mass. For simplicity, we 
assume a spatially constant, uniform accelerating gradient, $G$, 
even though the actual acceleration structure will consist of 
short, high acceleration regions interspersed with drift regions.
Refinement to include this detail should only change our smooth 
curves to segments of alternating flat and higher slope regions 
but with essentially the same average behavior. For convenience, 
we scale $G$ by the pion mass so that $g \equiv G/m_\pi$ has the 
dimensions of an inverse length,
\beq
	g m_\pi = dE_{\pi}/dx \ . 
\eeq

The energy of a pion as a function of distance along the 
accelerator is therefore
\beq
	E_{\pi}(x) = E_{I} + g m_\pi x \ .
\eeq
However, to follow the decay process, we will need the time rate 
of change of pion energy, viz.,   
\beq
	dE_{\pi}/dt  =  \frac{dE_{\pi}}{dx} \frac{dx}{dt} =  g m_\pi \beta(t) \ ,	
\eeq
where $\beta(t)$ is the pion velocity in units of $c$. Since 
$\gamma(t) = E_{\pi}/m_\pi$, we can rewrite Eq.\ (1) as
\beq
	d \gamma /dt = g \beta
\eeq
and since $\gamma =  1/\sqrt{1-\beta^2}$, 
\beq
	\frac{d\gamma}{dt} = \frac{\beta}{[1-{\beta}^2]^{\;3/2}} \frac{d\beta}{dt} \  .
\eeq
Eq.~(4) thus becomes
\beq
	\frac{1}{[1-{\beta}^2]^{\;3/2}}\frac{d\beta}{dt} = g \ . 
\eeq
This is easily integrated to obtain 
\beq
	\frac{\beta(t)}{[1-{\beta}^2(t)]^{1/2}} = g t 
		+ \frac{\beta(0)}{[1-{\beta}^2(0)]^{1/2}} \ ,
\eeq
or equivalently,  
\beq
	\beta(t) \gamma(t) =  g t + \beta(0) \gamma(0) \equiv g t + \alpha_0 \ . \label{eq:betagamma} 
\eeq
This allows us to identify
\bqry
	\beta(t) & = & \frac{g t +  \alpha_0}{\sqrt{1 + 
		({g t +  \alpha_0})^2}} \label{eq:betaoft} \ , \\
	\gamma(t) & = & \sqrt{1 + ({g t +  \alpha_0})^2} \ .
			\label{eq:gammaoft}
\eqry
Bear in mind that $\alpha_0$ is implicitly a function of the injected pion energy.
Eq.~(\ref{eq:betagamma}) is just the well-known result that a constant 
accelerating energy gradient causes the momentum to 
increase linearly with time, i.e., $p(t) = m_\pi g t + p_{I}$, 
where $p_{I} = m_\pi \alpha_0$ is the pion momentum at injection into the 
accelerator.

\subsection{Decay}

With $\gamma(t)$ in hand, we can solve for the pion 
decay function in the rest frame of the accelerator, viz. 
\beq
	\frac{dN_{\pi}}{dt}  =  - \frac{ \Gamma }{\gamma(t)} N_{\pi}(t)
	 =  - \; \frac{ \Gamma }{\sqrt{1 + (g t +  \alpha_0)^2}} N_{\pi}(t) \ ,
\eeq
where $\Gamma$ is the decay rate of the pion in its own 
rest frame. Integration of this equation yields 
\beq
	N_{\pi}(t) = N_{\pi}(0) \left[ \frac{(g t +  \alpha_0) + 
		\sqrt{1 +(g t +  \alpha_0)^2}}
	{ \alpha_0 + \sqrt{1 + \alpha_0^2}} \right]^{-\Gamma/g} \ .
\eeq
Since the number of neutrinos is just equal to the 
number of pions that have decayed, $N_{\nu}(t) = N_{\pi}(0) 
- N_{\pi}(t)$, it follows immediately that 
\beq
N_{\nu}(t) = N_{\pi}(0) \left( 1 - \left[ \frac{ \alpha_0 + 
	\sqrt{1 + \alpha_0^2}} {(g t +  \alpha_0) + \sqrt{1 + (g t +  \alpha_0)^2}}
	\right]^{\Gamma/g} \right) \ .	\label{eq:Nnuoft}
\eeq

\subsection{Distance Dependence}

We next convert this time dependence into spatial dependence, 
as a function of distance from the input of the linear accelerator. 
We can integrate $\beta = dx/dt$ given by Eq.\ (\ref{eq:betaoft}) to obtain
\beq
	x(t)  =  \int_{0}^{t} dt^{\prime} \beta(t^{\prime})
	 =  \frac{\sqrt{1+(gt+ \alpha_0)^{2}} - \sqrt{1+ \alpha_0^{2}}}{g} 
		\  , \label{eq:xoft}
\eeq
where we chose $x(0) = 0$.
Eq.\ (\ref{eq:xoft}) can easily be inverted to obtain
\beq
	t(x) = \frac{[(gx+\gamma(0))^2-1]^{1/2} - \alpha_0}{g c} \ , \label{eq:tofx}
\eeq
and this may be substituted into Eq.\ (\ref{eq:Nnuoft}) to obtain $N_{\nu}(x)$. 
Our result for the neutrino flux at a distance $x$ is, therefore,
\beq
	N_{\nu}(x) = N_{\pi}(0) \left[1 -  
		\left( \frac{ \alpha_0+\sqrt{1+ \alpha_0^2}}
		{gx + \sqrt{1+ \alpha_0^2} + [(gx + \sqrt{1+ \alpha_0^2}\;)^2 - 1 \; ]^{1/2} }
		\right)^{\Gamma/g} \right] .
\eeq
Noting that $m_\pi \sqrt{1+ \alpha_0^2} = E_{I}$, we convert this back to 
the original variables, 
\bqry
	N_{\nu}(x) & = & N_{\pi}(0) \left[ 1 - \left( \frac{p_{I}+E_{I}}
		{g m_\pi x + E_{I} + \sqrt{[g m_\pi x + E_{I}]^{2} - m_\pi^2}} 
		\right)^{\Gamma /g} \right] \nonumber \\
	& = & N_{\pi}(0) \left[ 1- \left( \frac{E_{I}+p_{I}}{E_x + p_x}
		\right)^{m\Gamma /G} \right] \label{eq:Nnuofx} \ ,
\eqry
where $E_{x}$ and $p_{x}$ are the energy and momentum of the 
pion after travelling a distance $x$ through the accelerator. 

\begin{figure*}
\begin{center} 
\includegraphics[ height=.8\textwidth , width=1.0\textwidth]{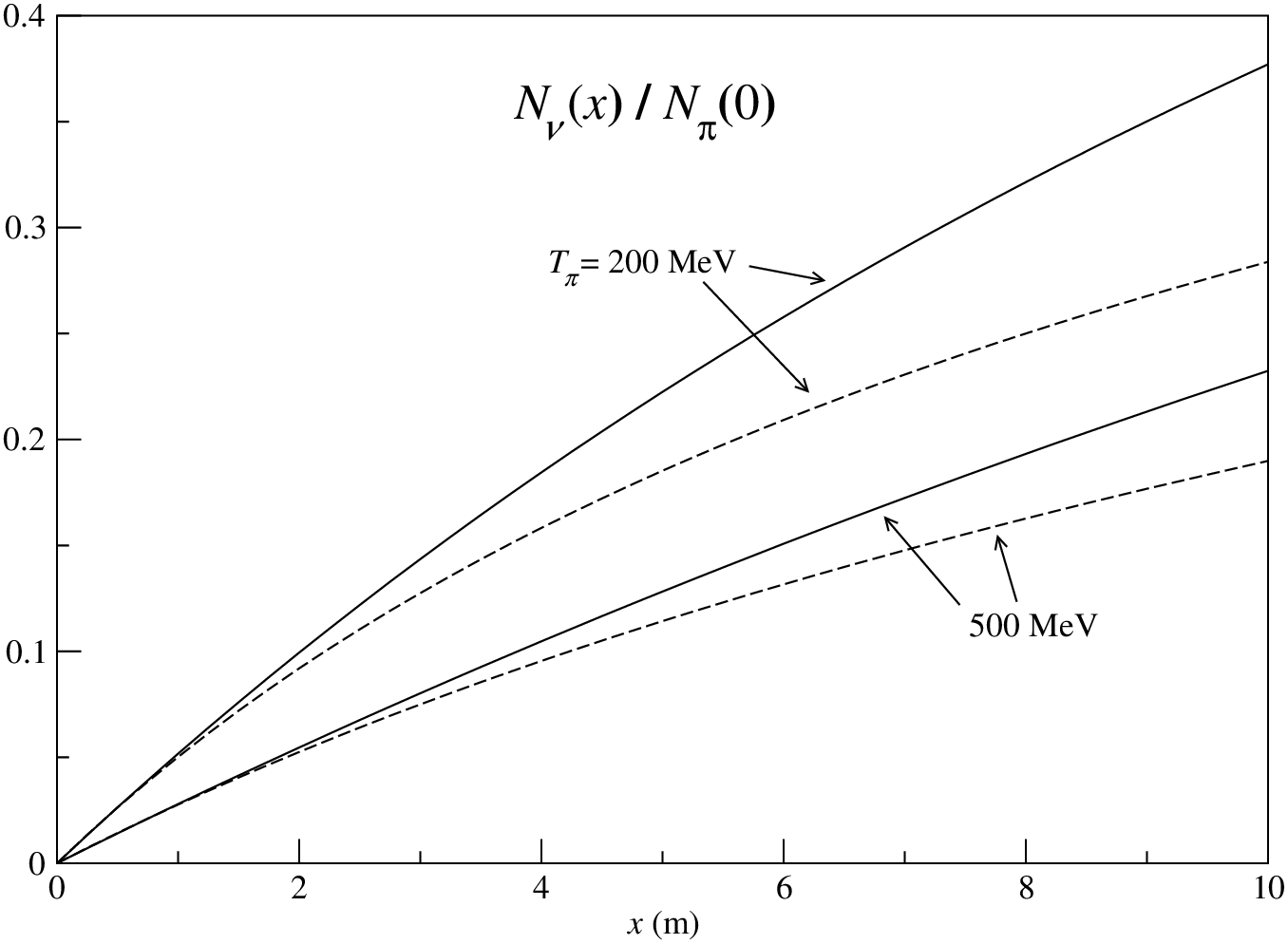} 
\end{center}
\caption{Dependence of $N_\nu(x)$, normalized by the number of input pions, 
$N_\pi(0)$, as a function of distance along the pion accelerator, $x$, 
for two pion kinetic energies spanning the production peak at injection 
into the accelerator. 
The solid curves are for a gradient of 10 MeV/m, while the dashed curves
are for 50 MeV/m.}
\label{fig:Nofx} 
\end{figure*}

Fig.\ \ref{fig:Nofx} shows the behavior of $N_\nu(x)$ for 
two different accelerating gradients and for two different initial 
pion kinetic energies (in the laboratory).
As expected, the $x$-dependence becomes more and more linear as the
injection energy increases.
Also, the higher the gradient, the flatter the curves become.
The reason for this is that a bigger gradient means higher energy pions, 
which have longer lifetimes (in the laboratory), whence fewer decays 
into neutrinos during acceleration.

The accelerator designer may therefore want to optimize between the
number of neutrinos showing up in the detector (normalized 
by the number of pions captured in the accelerator) and their
ultimate energies.
One might also want to consider the possibility of making a pion
{\em de}-celerator, if it turns out that lower energy neutrinos
have desirable properties for some experiments.

\subsection{Neutrino Energy Spectrum and Angular Divergence}

The neutrino energies will range over a wide spectrum, but for 
any specific time from injection of the pion bunch and angle of 
divergence from the beam axis of a neutrino scattering event 
in the detector, only one specific neutrino energy is possible. 

The momentum and energy, $E_{0}$, of a neutrino emitted in 
the pion rest frame is 
\beq
	E_{0}  =  \frac{m_{\pi}^{2}-m_{\mu}^{2}}{2m_{\pi}} =  29.8 {\rm\ MeV}
		\label{eq:Enurest} \ .
\eeq
For a decay where the neutrino is emitted at a polar angle $\theta_{R}$ (in 
the rest frame) with respect to the pion beam axis, the momentum,  
polar angle $\theta_{L}$, and energy $E_\nu$ in the lab frame are given by
\bqry
	 {\bf p} &=&  
		E_{0}\;[\gamma(\beta + \cos\theta_R), \; \sin\theta_R] \label{eq:pnulab} \ ,\\ 
	\tan\theta_{L}  &=& 
		\sin\theta_R/[\gamma(\beta + \cos\theta_R)] \label{eq:tanthetaL} \ ,\\ 
	 E_\nu &=&  E_{0}\;\sqrt{\sin^2\!\theta_R + \gamma^2(\beta + \cos\theta_R)^2} \nonumber \\
           &=& \gamma(1 + \beta\cos\theta_R) E_0
		\ , \label{eq:labEnu}
\eqry
where we show only the longitudinal and transverse momentum components, 
ignoring the azimuthal angle since the distribution in that angle is uniform. 

From Eq.\ (\ref{eq:labEnu}), neutrinos emitted exactly backward from the beam direction 
will have lab energies and momenta of $\gamma(1-\beta)\; E_{0}$, i.e., 
approximately zero for the most relativistic pions. Those emitted 
precisely forward will have lab energies and momenta of $\gamma(1+\beta)\; E_{0}$. 
For neutrinos emitted transverse (in the pion rest frame) to the beam direction, 
the lab energy and momentum will be $\gamma E_{0}$. 

From Eq.\ (\ref{eq:tanthetaL}), the singularity at $\cos\theta_R = -\beta$
corresponds to a laboratory angle of 90$^\circ$.
Neutrinos at this angle come from a very backward decay and have a
low energy of $E_0/\gamma$.
Those that come from even more backward decays have yet lower energies.
The maximum laboratory angle is 180$^\circ$.
That is, $\theta_L$ as a function of $\theta_R$ rises monotonically from 
0 to 180$^\circ$, and does so more steeply in the backward $\theta_R$ direction.

Neutrinos are emitted between the two extreme energies along a curve in 
$E_\nu$-$\theta_{L}$ space. 
To make such a plot, one needs to find the value of $\cos\theta_R$
corresponding to a given laboratory angle, $\theta_L$.
From the inverse Lorentz transformation, from the laboratory to the pion rest frame,
and with some algebra,
\beq
	\cos\theta_R = \cos\left[\tan^{-1}\left(
		\frac{\sin\theta_L} {\gamma (\cos\theta_L - \beta)} \right)\right]
		= \frac{\cos\theta_L - \beta}{1 - \beta\cos\theta_L}
	\ . \label{eq:costhetaR}
\eeq

\begin{figure*}
\begin{center}
\includegraphics[height=0.6\textwidth , width=0.9\textwidth ]{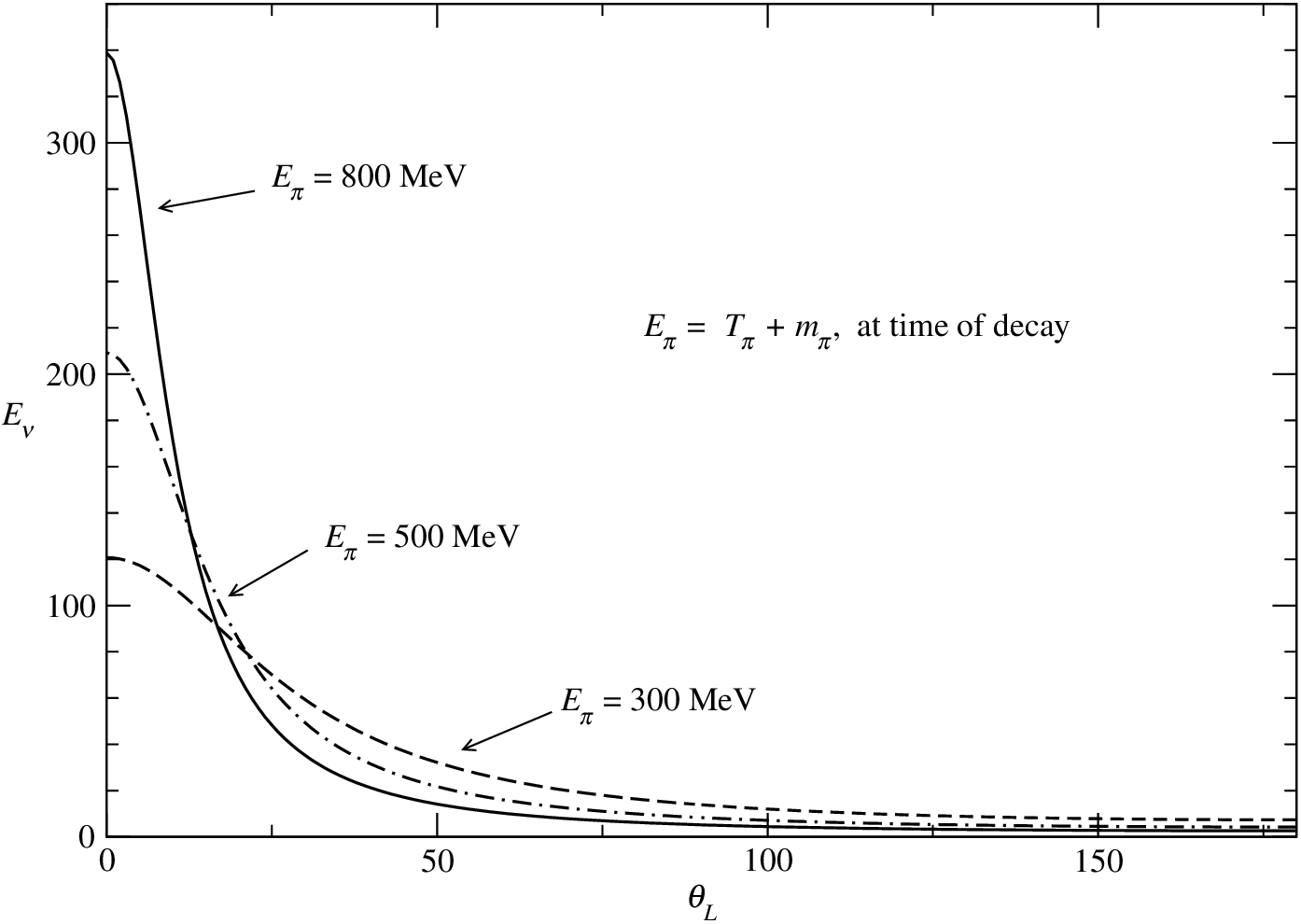}
\end{center}
\caption{The neutrino energy, $E_\nu$ (in MeV), as a function of laboratory polar 
angle $\theta_L$ from decay in the acceleration section at a given time from 
bunch injection (or position in the accelerator)
for three pion kinetic energies at the time of decay. 
Each pion energy corresponds to a specific location for the decay in the
accelerator section.}
\label{spectrum} 
\end{figure*}

Having found the value of $\cos\theta_R$ for a given (or derived) $\theta_L$,
the neutrino energy $E_\nu$ is then given by Eq.~(\ref{eq:labEnu}),
or more directly, $E_\nu = (1 - \beta\cos\theta_L)^{-1} E_0 / \gamma$.
Figure \ref{spectrum} shows curves of the neutrino energy as a function of 
$\theta_L$ for three different (total) energies of the pion at the time of its decay.
Note that, for the higher pion energies, the neutrino energy $E_\nu$ is
more forward peaked and, of course, is larger there.
Taking into account that neutrino cross sections are proportional to $E_\nu^{\ 2}$
at low $E_\nu$, this suggests that, for economy, the neutrino detector could well 
be long and narrow, subtending a small solid angle along the beam direction.

\subsection{Determining $E_\nu$ From Event Timing}

Fig.\ 3 depicts an example of what we have in mind, a pion accelerator 
which is 10 m long.\cite{ftn2}
This is followed by an air-filled drift space (i.e., a decay volume) of 14 m,
a 10 m long beam dump, such as earth or iron, and the neutrino detector, 
presumed, for cost-effectiveness, to be a long cylinder aligned with the 
beam axis.\cite{ftn3}
The drift space could also provide space for experiments using the 
pions which emerge from the accelerator.
It also might contain bending magnets to deflect charged particles
from the beam axis.\cite{ftn4}

Neutrinos all travel at velocity $c$, which is discernably larger than 
the velocity of the pions (up to $\approx$ 1 GeV pion energy).
Thus the time of scattering and the position of the event in the detector 
are uniquely correlated (up to resolution limits) 
with the energy of the neutrino.
However, low-energy neutrino cross sections vary as $E_\nu^2$,
so most of the neutrino events will come at forward angles less 
than $\approx 30^\circ$.

\begin{figure*}
\begin{center}
  \rotatebox{0}{\includegraphics[ height=0.6\textwidth , width=0.9\textwidth]
	{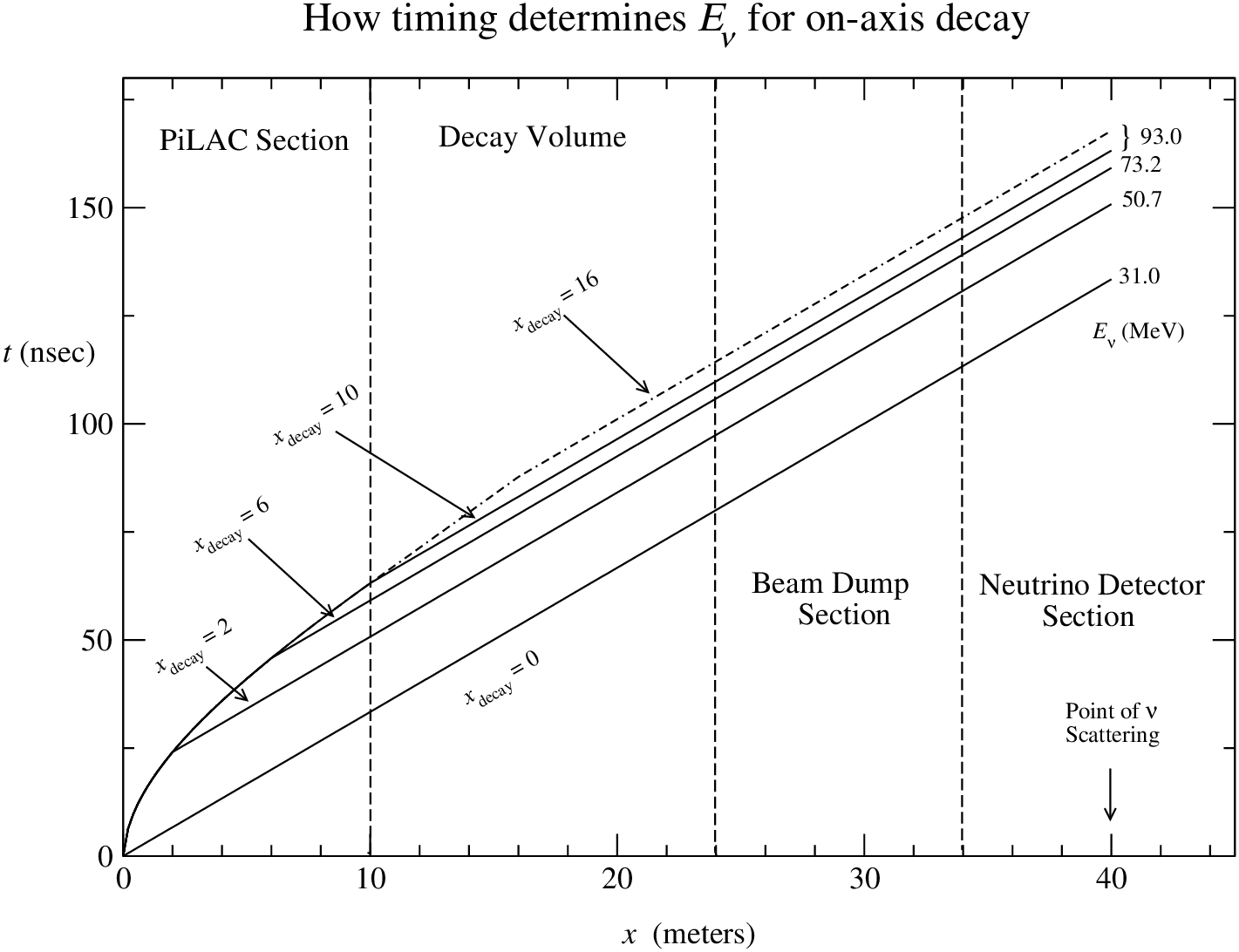} }
\end{center}
\caption{
Relation between decay time after pion bunch injection and time of 
scattering event in detector for {\bf on-axis} $\nu$'s of the highest energy.
This is for an {\em unrealistic} injection energy of $T_\pi$ = 100 keV, so that
the curvatures before decay are highly exaggerated.
} 
\label{ETcorrln} 
\end{figure*}

We discuss first the case for those higher-energy neutrinos emitted at 
$0^\circ$ along the beam direction.
Fig.\ \ref{ETcorrln} shows how the change in pion velocity along the 
accelerator, followed by speed-of-light propagation of the neutrino 
from the decay point, determines the neutrino energy.
Note that this figure is for an {\em unrealistic} injection energy of 
$T_\pi$ = 100 keV, simply to show the principle involved.\cite{ftn5}
The idea is to measure the difference between the time when the neutrino scattering 
event occurs and that for a light-like signal that begins at the time
the pions are injected into the pion accelerator.
The neutrino energy $E_\nu$ in this on-axis case is given by
\beq
	E_{\nu,\;{\rm on-axis}} = \gamma(t_{\rm dec})\;[1 + \beta(t_{\rm dec})]\;E_0
		\ , \label{eq:Enuonaxis} 
\eeq
where $t_{\rm dec} = t(x_{\rm dec})$ is given by Eq.\ (\ref{eq:tofx}) 
and $E_0$ by Eq.\ (\ref{eq:Enurest}).
The value of $x_{\rm dec}$ needed for $t_{\rm dec}$ is obtained by solving 
the more general expression for the event time,
Eq.\ (\ref{eq:tevent}), given below.

Note also that some pions will decay after leaving the accelerator.  
In such cases, as in the dot-dashed curve in Fig.\ 3, 
the timing for the event will be later than
the case shown when $x_{\rm decay}$ = 10 m, but $E_\nu$ will be the same,
i.e., 93.0 MeV in this example.

For {\em more realistic} injection energies, e.g., $T_\pi$ = 350 MeV,
the curves in the pion accelerator section are much, much flatter
than those shown in Fig.\ \ref{ETcorrln}.
The time differences are quite small but technically feasible.
\begin{table}[b] 	
\begin{center}
\begin{tabular}{||c|c|c|c|c||} \hline
$x_{\rm decay}$ (m) & \multicolumn{2}{c|}{$G$ = 10 MeV/m} & 
		\multicolumn{2}{c||}{$G$ = 50 MeV/m} \\ \hline
    &  $\Delta t$ (nsec) & $E_\nu$ (MeV)
	&  $\Delta t$ (nsec) & $E_\nu$ (MeV) \\ \hline
0.0  & 0.000 & 205 & 0.000 & 205 \\
2.0  & 0.275 & 214 & 0.236 & 249 \\
4.0  & 0.528 & 223 & 0.401 & 292 \\
6.0  & 0.762 & 231 & 0.522 & 335 \\
8.0  & 0.978 & 240 & 0.616 & 379 \\
10.0 & 1.179 & 249 & 0.691 & 422 \\ \hline
\end{tabular}
\end{center}
\caption{The calculations shown here are for $T_\pi$ = 350 MeV at injection into PILAC.}
\end{table}
For an injection energy of $T_\pi$ = 350 MeV, Table I shows,
for two choices of accelerating gradient $G$, the time differences, in nanoseconds, 
and the higher-energy neutrinos' associated energies.
Note that, to distinguish the two highest energies shown for $G =$ 50 MeV/m, 
one must have timing resolution of about 50 picoseconds.

A similar but more complicated relation occurs for radially off-axis points, 
where the decay angle must be included in the correlation. 
(See Fig.\ \ref{spectrum}.)
The procedure for determining $E_\nu$ from the time the event occurs
(relative to the time of injection of the pions into the accelerator) is 
as follows.

If the position of the event in the detector is given by 
$(x_{\rm evt}, \rho_{\rm evt})$, where $x_{\rm evt}$ 
is measured from the
injection point and $\rho_{\rm evt}$ is the transverse distance from the
beam axis, the time of the event is
\beq
	t_{\rm evt}(x_{\rm dec},x_{\rm evt},\rho_{\rm evt}) =
		t_{\rm dec}(x_{\rm dec}) + t_\nu (x_{\rm evt} - x_{\rm dec}, \rho_{\rm evt})
		\ , \label{eq:tevent}
\eeq
where $t_{\rm dec}(x)$ is given by Eq.\ (\ref{eq:tofx}) and
$t_\nu (x, \rho) = \sqrt{x^2 + \rho^2}/c$.
Equation (\ref{eq:tevent}) can be solved numerically for $x_{\rm dec}$.
The implicit dependence in $t_{\rm dec}$ on the gradient, $G$,
and the kinetic energy $T_\pi$ of the injected pions should be kept in mind.

Given the extracted value for $x_{\rm dec}$, we can infer that the neutrino 
was emitted at a laboratory angle of
\beq
	\theta_L = \tan^{-1} [\rho_{\rm evt}/(x_{\rm evt} - x_{\rm dec})] \ . 
	\label{eq:thetalab}
\eeq
The pion energy at the time it decays is also determined by $x_{\rm dec}$.
From Eqs. (\ref{eq:gammaoft}) and (\ref{eq:tofx}),
\bqry
	\gamma_{\rm dec} &=& \sqrt{1 + [gt(x_{\rm dec}) + \alpha_0]^2}
		\label{eq:gammadec} \ , \\
	\beta_{\rm dec} &=& \sqrt{\gamma_{\rm dec}^2 - 1}\; /\gamma_{\rm dec}
		\label{eq:betadec} \ , \\
	E_\pi &=& m_\pi \; \gamma_{\rm dec} \ . \label{eq:Epidec}
\eqry

From the values of $\theta_L$, $\beta_{\rm dec}$, and $\gamma_{\rm dec}$ 
the cosine of the decay angle in the pion rest frame
is then found by substituting these values into Eq.\ (\ref{eq:costhetaR}):
\beq
	\cos\theta_R = 
	\frac{\cos\theta_L -\beta_{\rm dec}}{1 - \beta_{\rm dec}\cos\theta_L}
		\ . \label{eq:costhetaRdecay}	
\eeq
The neutrino energy $E_\nu$ is then determined 
by substituting this value of $\cos\theta_R$ into Eq.\ (\ref{eq:labEnu}).

\begin{table}	
\begin{center}
\begin{tabular}{||c|c|c|c|c|c|c|c||} \hline
   \multicolumn{2}{||c|}{ } & \multicolumn{3}{c|}{$G$ = 10 MeV/m} & 
		\multicolumn{3}{c||}{$G$ = 50 MeV/m} \\ \hline
$\Delta t$ (nsec) & $\ \rho_{\rm evt} \ $ 
	& $x_{\rm dec}$ & $\ \theta_L \ $ & $E_\nu$ 
	& $x_{\rm dec}$ & $\ \theta_L \ $ & $E_\nu$ \\ \hline
0.250 & 0.3   & 1.81 &  7.9  & 213    & 2.15 &  7.9 & 252 \\
      & 0.6   & 1.72 & 15.7  & 213    & 2.02 & 15.8 & 249 \\
      & 0.9   & 1.58 & 23.4  & 212    & 1.82 & 23.6 & 245 \\ 
      \hline
0.500 & 0.3   & 3.77 &  8.3  & 222    & 5.58 &  8.7 & 327 \\
      & 0.6   & 3.67 & 16.5  & 221    & 5.35 & 17.3 & 322 \\
      & 0.9   & 3.50 & 24.7  & 221    & 4.99 & 25.7 & 314 \\ 
      \hline
0.750 & 0.3   & 5.89 &  8.9  & 231    & 11.9 & 10.7 & 463 \\
      & 0.6   & 5.77 & 17.5  & 231    & 11.3 & 20.9 & 450 \\
      & 0.9   & 5.58 & 26.1  & 230    & 10.5 & 30.6 & 433 \\ 
      \hline
\end{tabular}
\end{center}
\caption{ 	
Extractions of the decay point $x_{\rm dec}$ (in m),
neutrino laboratory angle $\theta_L$ (in mrad),
and $E_\nu$ (in MeV) for a scattering event
taking place at various scattering times $\Delta t$ in the detector 
at three values of off-axis radial position $\rho_{\rm evt}$.
This case is for an event occurring in the detector at $x_{\rm evt}$ = 40.0 m 
from the accelerator injection point.
We have assumed here that $T_\pi$ = 350 MeV at 
injection into PILAC, as in Table I. }
\end{table}
Table II shows the correlations between measured scattering times $\Delta t$
and off-axis event positions $\rho_{\rm evt}$ and the extracted values for
the decay point $x_{\rm dec}$, laboratory neutrino angle $\theta_L$, and $E_\nu$, 
for the same two accelerating gradients and injection energy used in Table I.
Because of the small solid angle subtended by the detector, in this example,
the time differences at off-axis points are not much 
different from those shown in Table I.
Thus transverse spatial resolution becomes a significant contributor to 
determination of the neutrino energy, implicitly demanding a fine-grained detector.

Fig.\ 4 illustrates the correlation for an off-axis detection of a 
neutrino between the time of the event and the pion decay point and angle of 
decay (as seen in the lab frame), whence the energy of the reacting neutrino.

\section{Discussion} 

Note that, as shown by Eq.\ (\ref{eq:Nnuofx}), the important scale for the 
acceleration gradient is set by the lifetime of the pion in units of pion 
mass, namely, 
\beq
G = m_\pi \Gamma_{\pi}/c = 17.9 \;   {\rm MeV/m} \ . 
\eeq
Gradients exceeding this value are now routinely available, 
which makes possible very tightly focussed neutrino beams and 
unusual neutrino energy spectral distributions. 

The higher the gradient can be made, the more pions survive to 
the exit energy and the neutrino spectrum becomes more and more 
weighted to higher energies. The higher energy component of the 
neutrino beam also becomes more focused, projecting a smaller 
image on the target. 

The pion particle flux expected from the design in Ref.\ \cite{PILAC} was 
$10^{9} s^{-1}$ at the target end of the accelerator. With the gradient 
given above, the accelerator would be on the order of 5 lifetimes in 
length. Thus, while the neutrino flux at the highest energy available 
would be on the order of $10^{9} s^{-1}$, the total flux over the entire 
spectrum would be on the order of $10^{11} s^{-1}$ (since nearly every 
injected pion will decay into a neutrino). This certainly presents 
a challenge to experimenters to design detectors with a useful event rate, 
as the flux is not as large as produced by simple beam target-decay volume 
sources. 

An additional order of magnitude in flux should be available if the PILAC 
system were to be mounted at the Spallation Neutron Source at Oak Ridge 
National Laboratory. 

We thank J.~Jenkins, W.~Louis, G.~Mills, and C.\ Morris for valuable conversations.
This work was carried out under the auspices of the National Nuclear Security 
Administration of the U.S. Department of Energy at Los Alamos National Laboratory 
under Contract No. DE-AC52-06NA25396

\newpage

\begin{figure*}
\includegraphics[height=0.6\textwidth , width=0.9\textwidth]{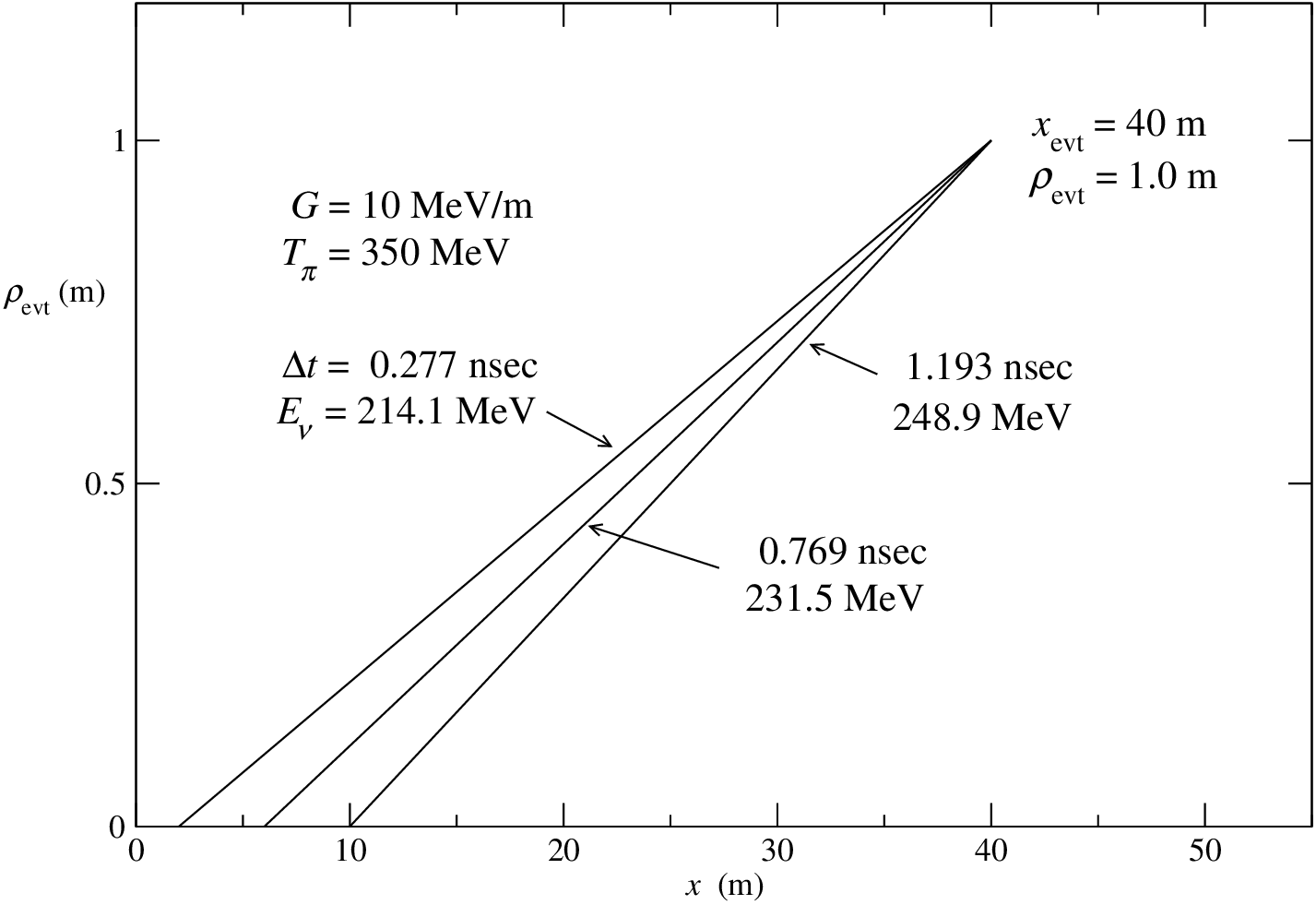} 
\caption{Relation between the reacting neutrino's energy, $E_\nu$, and the 
time difference between the time of bunch injection 
and the scattering event time in the detector 
for a particular event position, here at 40 meters from the start of the pion 
accelerator and at an off-axis distance of 1 meter.
It shows the differences in $E_\nu$ and time difference $\Delta t$ for 
three different pion decay positions.
The kinetic energy of pions at injection has been taken to be 350 MeV and 
the accelerating gradient is 10 MeV/m.
Thus, in this example, if the absolute time of this scattering event can be 
determined relative to the injection time with an accuracy of better than $\sim 
500$ ps, then the energy of the incident neutrino that initiated the event will be 
determined to an accuracy of better than $\approx 15$ MeV. } 
\label{AngleEnuCorreln} 
\end{figure*} 


\begin{thebibliography}{99}
\bibitem{ftn1}
{In the examples given in the tables below, we will use two values of gradients,
10 MeV/m and 50 MeV/m.  The latter value is rather speculative, but might
be achievable in the future.  In this regard, for a F0D0 linac rather than the 
continuous gradient design we assume here (for simplicity of presentation),
the average gradient will be less than that of a given RF cavity.  Also, the 
gradient achievable could depend upon on aperture required to match the 
emittance of the pion beam being accelerated.  Our only intention, in showing 
the 50 MeV/m case, is to indicate the effects of achieving high gradients.  As 
noted in the introduction, even at lower gradients, the design has a significant 
value in affording the determination of the energy of the incident neutrino event 
by event.}
\bibitem{ftn2}
{The PILAC design \cite{PILAC} was for a rather longer
accelerator, 102.9 m, but this was to get the highest possible energies 
for the pions.  Also, the accelerating gradient in that design
was only 12.5 MeV/m, befitting the technology available at that time.}
\bibitem{ftn3}
{Any realistic detector proposal could only be made after a realistic 
accelerator design has been developed.  Our hope is that a highly 
instrumented device, long but with a relatively small transverse size, 
might be feasible.}
\bibitem{ftn4}
{Without such a bend, muons from pion decays might otherwise penetrate
the beam dump and give background events in the neutrino detector which
follows.  The background from neutrons in the beam direction should be
small.  Note that the injection ``snake'' severely limits the pions
in the linac in both momentum and charge, so opposite-sign pions and
muons should be totally negligible.}
\bibitem{ftn5}
{Details dependent on realistic beam dispersion and timing errors will 
blur these lines.}
\end{thebibliography}
\end{document}